\documentclass[%
 reprint,
superscriptaddress,
 amsmath,amssymb,
prb,
floatfix,
]{revtex4-2}

\usepackage{graphicx}% Include figure files
\usepackage{subfigure}
\usepackage{dcolumn}% Align table columns on decimal point
\usepackage{bm}% bold math
\usepackage{dsfont}
\usepackage{xcolor}
\usepackage{placeins}
\usepackage{tikz}
\definecolor{C0}{HTML}{1F77B4}
\definecolor{C1}{HTML}{FF7F0E}
\definecolor{C2}{HTML}{2CA02C}
\definecolor{C3}{HTML}{D62728}
\definecolor{C4}{HTML}{9467BD}
\definecolor{C5}{HTML}{8C564B}
\definecolor{C6}{HTML}{E377C2}
\definecolor{C7}{HTML}{7F7F7f}
\definecolor{C8}{HTML}{BCBD22}
\definecolor{C9}{HTML}{17BECF}

\begin{document}

\preprint{APS/123-QED}

\title{
%A scalable many-body quantum simulation workflow:\\demonstration on a Hubbard molecule
%Solving a ring-like fermionic Hubbard model on a %current 
%quantum computer
Simulating a ring-like Hubbard system with a quantum computer
}

\author{Philippe Suchsland}
 \affiliation{%
Institute for Theoretical Physics, ETH Zurich, 8093 Zurich, Switzerland
 }%
  \affiliation{%
 IBM Quantum, IBM Research -- Zurich, 8803 Rueschlikon, Switzerland
}%
 \affiliation{%
 Department of Physics, University of Zurich, Winterthurerstrasse 190, 8057 Zurich, Switzerland
 }%
\author{Panagiotis Kl. Barkoutsos}
\affiliation{%
 IBM Quantum, IBM Research -- Zurich, 8803 Rueschlikon, Switzerland
}%
\author{Ivano Tavernelli}
\affiliation{%
 IBM Quantum, IBM Research -- Zurich, 8803 Rueschlikon, Switzerland
}%
\author{Mark H Fischer}%
\affiliation{%
 Department of Physics, University of Zurich, Winterthurerstrasse 190, 8057 Zurich, Switzerland
}%
\author{Titus Neupert}
\affiliation{%
 Department of Physics, University of Zurich, Winterthurerstrasse 190, 8057 Zurich, Switzerland
}%

\date{\today}

\begin{abstract}
We develop a workflow to use current quantum computing hardware for solving quantum many-body problems, using the example of the fermionic Hubbard model. 
Concretely, we study a four-site Hubbard ring that exhibits a transition from a product state to an intrinsically interacting ground state as hopping amplitudes are changed.
We locate this transition and solve for the ground state energy with high quantitative accuracy using a variational quantum algorithm executed on an IBM quantum computer.
Our results are enabled by a variational ansatz that takes full advantage of the maximal set of commuting $\mathbb{Z}_2$ symmetries of the problem and a Lanczos-inspired error mitigation algorithm. They are a benchmark on the way to exploiting near term quantum simulators for quantum many-body problems.
\end{abstract}

\maketitle

%\tableofcontents

%\textit{Introduction.}
Fully programmable quantum computing devices are an emerging technology for which a series of milestones have been demonstrated over the past years leading to devices with lower error rates and tens of qubits~\cite{progress_ibm,state_of_play, Wei2020}.
Despite being noisy, these devices cater to a range of envisioned applications, including quantum search~\cite{Jones1998}, quantum machine learning~\cite{Biamonte2017, Schuld2019, spsagood} and finance~\cite{Woerner2019, Stamatopoulos2020}. 

An application for which quantum computers are innately advantageous are quantum many-body problems that arise in condensed matter physics~\cite{phase_transition1,phase_transition2,phase_transition3,dtc_on_qc,many_body_quantum} and quantum chemistry~\cite{efficent_vqe,vqe_google}. 
The classical computational cost to investigate such systems grows exponentially with the system size, often exceeding hardware limitations before the behavior of a thermodynamically large system can be deduced. 

Current fully programmable quantum computing devices are limited by the decoherence times of the qubits and gate as well as readout errors. IBM's hardware, a representative industry standard, reaches about 100~$\mu$s decoherence time and a few percent readout error. Critical are the two-qubit gates with an error of about 1\% and operation times of 0.2--0.5~$\mu$s. %One-qubit gates are comparably accurate and fast. 
This limits the number of two-qubit gates available for algorithms with quantitative accuracy to about 20 and with it the number of qubits that can be entangled. Consequently, quantum computation of many-body ground states with high quantitative accuracy has only involved 2--3 qubits up to date~\cite{prev_work_11,error3,prev_work_13,vqe_google,Note1}\footnotetext[1]{Other works used more qubits, but did not reach the same level of accuracy~\cite{efficent_vqe,error3}.}.
In turn, \emph{classical simulations} of such quantum 
algorithms suggest that the number of required two-qubit gates rises steeply with Hilbert space size, requiring $O(100)$ two-qubit gates already for a four-site (spinful fermionic) Hubbard model, putting it out of reach for current quantum computing hardware~\cite{hubbard_sim_qc,prev_work_32,prev_work_33,hubbard_sim_qc3,choquette2020quantum}.

Here, 
we push these boundaries and establish a scalable workflow for solving prototypical strongly-correlated quantum many-body problems  on current quantum computers. Specifically, we focus on the iconic fermionic Hubbard model, which we investigate on a ring with four sites. Conceived to resolve the puzzles of high-temperature superconductivity, the solution to the Hubbard model rose to become a question of scientific value on its own right~\cite{hubbard_sim_qc,hubbard_sim_qc2,hubbard_sim_qc3}. Concretely, the four-site Hubbard ring shows a transition between a ground state that is adiabatically connected to a single Slater determinant and a ground state that is intrinsically interacting as long as time-reversal and rotation symmetries of the ring are respected. The latter state is a building block for a two-dimensional quantum phase called fragile Mott insulator, a symmetry-protected topological phase, when rings are connected into an extended square lattice~\cite{fmi}. 

To solve for the ground state of the Hamiltonian, we employ a hybrid quantum-classical variational algorithm. All measurements of quantum-mechanical expectation values are performed on an IBM quantum computer, while optimization steps are performed classically. 

As we show in the following, we obtain ground-state energies with an accuracy of a few percent in units of the typical energy scales of the Hamiltonian. Three main theoretical advances are combined into our workflow:
(i) We introduce a variational form for the ground state, called \emph{adaptive $R_yR_z$ ansatz}%~\cite{XX}\textcolor{red}{here we need a citation (philippe:) what kind? we introduce/invent it}
, that strikes a balance between the number of two-qubit gates and variational parameters, which is optimal for the performance characteristics of the quantum device. (ii) We fully exploit the symmetries of the system through \emph{tapering off} of qubits~\cite{taper}. This allows us to also track the ground state transition of the Hubbard ring more precisely. (iii) To reduce  systematic errors, we employ a recently introduced Lanczos-inspired mitigation algorithm~\cite{suchsland2020algorithmic}. %, which we recently introduced~[].

%Scheme to implement time evolution Hubbard model proposed 1997 \cite{early_prop_hubbard}.

%Transition of Pollmann\cite{phase_transition1}

%Transition similar to Pollmann\cite{phase_transition2}

%Transition in Majoranas\cite{phase_transition3}

\section{Results and Discussion}
\begin{figure}[t]
\centering
%\subfigure{\input{figures/molecule}}
\subfigure{\begin{tikzpicture}[xscale=0.7,yscale=0.7]                   

% \chemsetup[orbital]{
%     overlay ,
%     opacity = 1 ,
%     p/scale = 1.6 ,
%     p/color = C2 ,
%     s/color = C2 , %changing to red!50 is no help%
%     s/scale = 1.2
%   }
%   \def\sorb{ \chemfig{\orbital{s}} }

%\usepackage{chemfig,chemmacros}
%\chemsetup{modules=all}

\node at (-7.2,{1.5}) {(a)};

\def\d{1.7}
\node (A) at (-6-0.5,{-\d/2-0.2}) {};
\node (B) at (\d-6-0.5,{-\d/2-0.2}) {};
\node (C) at (\d-6-0.5,{\d/2-0.2}) {};
\node (D) at (0-6-0.5,{\d/2-0.2}) {};

\draw[line width=0.5mm]  (A) to (B);
\draw[line width=0.5mm]  (B) to (C);
\draw[line width=0.5mm]  (C) to (D);
\draw[line width=0.5mm]  (D) to (A);

%\foreach \v in {A,B,C,D}
%    { \node at (\v) {\sorb};}

%is exactly the same like chemfig just without chemfig
\foreach \v in {A,B,C,D}
    { \shade[ball color = C2] (\v) circle (0.4);}

%\node[above left=4] at (A) {\color{C2} 0.};
%\node[above left=4] at (D) {\color{C2} 3.};
%\node[above left=4] at (B) {\color{C2} 1.};
%\node[above left=4] at (C) {\color{C2} 2.};

\draw[line width=0.75mm,->,bend left,color=C1]  (B) to [out=30,in=150] (A);
\node at ({\d/2-6-0.5},{-0.6-\d/2-0.2}) {\large \color{C1}$t$};
\draw[line width=0.75mm,->,bend right,color=C0]  (C) to [out=-30,in=220] (A);
\node at ({\d/2-6-0.5},{0.4+\d-\d/2-0.2}) {\large \color{C0}$t'$};

\node at (-3.5,{1.5}) {(b)};

        \newcommand\st{3}    
        \newcommand\sty{1.4}                                                                     
        \draw[scale=1,domain=-\st:\st,smooth,variable=\x,C1,very thick,opacity=0.3] plot ({\x},{-cos(\x*180/\st)}); 
        \draw[scale=1,domain=-\st:\st,smooth,variable=\x,C0,very thick,opacity=0.3] plot ({\x},{cos(2*\x*180/\st)}); 
        \draw[->,very thick] (-\st,-1.4) -- (\st,-1.4) node[right] {$\boldsymbol \lambda$};                                       
        \draw[->,very thick] (-\st,-1.4) -- (-\st,1.4) node[above,right] {\color{C1}$E/t$,\color{C0}$E/t'$}; 
        \draw[gray] (-\st,0) node[left] {$0$} -- (\st,0) ; 
        
        \node[left,gray] at (-\st,-1) {$-2$}  ; 
        
        \foreach \i [count=\j] in {-\st,-\st/2,0,\st/2} {
            \draw[very thick] (\i,-\sty+0.15) -- (\i,-\sty-0.15);% node[below] {\j};
        }
        \draw[very thick,gray] (-\st-0.1,-1) -- (-\st+0.1,-1);
        \node[below] at (-\st,-\sty-0.1) {$1$};
        \node[below] at (-\st/2,-\sty-0.1) {$i$};
        \node[below] at (0,-\sty-0.1) {$-1$};
        \node[below] at (\st/2,-\sty-0.1) {$-i$};

        \draw [C1,line width=0.75mm] (-\st/2-0.3,{-cos(-180/2)}) -- (-\st/2+0.3,{-cos(-180/2)});  
        \draw [C1,line width=0.75mm] (\st/2-0.3,{-cos(-180/2)}) -- (\st/2+0.3,{-cos(-180/2)}); 
        \draw [C1,line width=0.75mm] (-0.3,{-cos(-180*0)}) -- (0.3,{-cos(-180*0)});
        \draw [C1,line width=0.75mm] (-\st-0.3,{-cos(-180)-0.03}) -- (-\st+0.3,{-cos(-180)-0.03});
        
        %\draw [C1,very thick] (2.5-0.,0.3) -- (2.5+0.4,0.3);
        %\draw [C1,very thick] (2.5+0.5,0.3) -- (2.5+0.9,0.3);
        %\draw [C1,very thick] (2.5-0.,-0.3) -- (2.5+0.4,-0.3);
        %\draw [C1,very thick] (2.5+0.5,-0.3) -- (2.5+0.9,-0.3);
        %\draw[thick] (2.5,-0.6) -- (3.5,-0.6) node[right] {$\lambda$}; 
        %\draw[thick] (2.7,-0.5) -- (2.7,-0.7);
        %\draw[thick] (3.2,-0.5) -- (3.2,-0.7);
        %\node[below] at (2.7,-0.7) {$i$};
        %\node[below] at (3.2,-0.7) {$-i$}; 
        
        \draw [C0,line width=0.75mm] (-\st/2-0.3,{cos(-180)}) -- (-\st/2+0.3,{cos(-180)});  
        \draw [C0,line width=0.75mm] (\st/2-0.3,{cos(-180)}) -- (\st/2+0.3,{cos(-180)});  
        \draw [C0,line width=0.75mm] (-0.3,{cos(-180*0*2)}) -- (0.3,{cos(-180*0*2)});
        \draw [C0,line width=0.75mm] (-\st-0.3,{cos(-180*2)+0.03}) -- (-\st+0.3,{cos(-180*2)+0.03});
        
        %\node[C1,left] at (2.5-0.,-0.3) {b)}; 
        %\node[C1,left] at (2.5-0.,0.3) {a)}; 
        \node[C1] at (0,{-cos(-180/\st*0)}) {\Large $\boldsymbol \uparrow \boldsymbol\downarrow$}; 
        \node[gray] at (\st/2,{-cos(-180/2)}) {\Large $\boldsymbol\uparrow \boldsymbol\downarrow$};
        %\node[C1] at (-1.25,{-cos(-180/\st*1.25)-0.3}) {\large $-$};
        \node[gray] at (-\st/2,{-cos(-180/2)}) {\Large $\boldsymbol\uparrow \boldsymbol\downarrow$};
        %\node[C1] at (2.5+0.7,0.3) {\large $\uparrow \downarrow$};
        %\node[C1] at (2.5+0.2,-0.3) {\large $\uparrow \downarrow$};
        %\node[C1] at (1.25,{-cos(-180/\st*1.25)+0.3}) {\large $-$};
        
        %\draw[rounded corners] (-1.25-0.25,{-cos(-180/\st*1.25)+0.15}) rectangle (1.5,0.45);
        
        %\draw[rounded corners] (-1.25-0.25,{-cos(-180/\st*1.25)-0.15}) rectangle (1.5,-0.45);
        
        \node[C0] at (\st/2,{cos(-180)}) {\Large $\boldsymbol\uparrow \boldsymbol\downarrow$}; 
        \node[C0] at (-\st/2,{cos(-180)}) {\Large $\boldsymbol\uparrow \boldsymbol\downarrow$};  
\end{tikzpicture}}
\caption[]{
%\textcolor{blue}{(Panos) This Figure looks like there is a lot of white space. Can we make the 4-site latttice a bit bigger and wider?} 
%\textcolor{red}{(ivano) the $\lambda$s are used to construstruct the 'symmetrized' states generated by the new $c$s in Eq. 3. However, the connection with the eigenvalues in this figures remains hidden. Can you add a formula for $E_{\lambda}$ in the text or in the SI? Is there a way to visualize the symmetrized states on the lattice? It could also help the interpretation.}
The four-site system is shown left in (a) alongside the energy levels for the nearest-neighbour and next-nearest-neighbour hoppings in (b). In (a) the (next-)nearest-neighbour hopping is visualized with an orange (blue) arrow and labelled with $t$ ($t'$).
In (b), the fermionic states of the four-site molecule are labeled by $(\lambda,\sigma)$. For half-filling with only next-nearest-neighbour hopping $t' \neq 0, t=0, E/t'=2\lambda^2$ the ground state is non-degenerate as shown in blue, while it is degenerate for nearest-neighbour hopping only $t\neq0,t'=0,E/t=\lambda+\lambda^*$, where the occupied states are orange and  half-occupied states are gray.
}
\label{fig:foursites}
\end{figure}
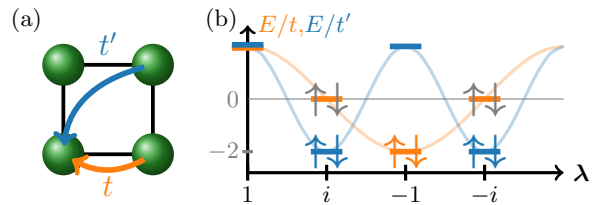

\textit{The model.} 
%
%
%{\color{C2}\textit{1.1 CI, BI, fMI.}} A crystalline %insulator is a system in which the ground state is separated from the first excited state by a finite energy gap. We call it a band insulator (BI) if it is adiabatically connected to a non-interacting system~\cite{fmi}. In contrast to BIs, Mott insulators are insulating phases due to the electron-electron interactions and become metals if interactions are turned off~\cite{mott}. These can be extended to fragile Mott insulators (FMIs), which additionally transform non-trivially under the operations of the point group of the crystal~\cite{fmi}. They are only distinct from BI if the symmetries are not broken and, hence, are called fragile. With symmetries present FMIs are a distinct phase of matter since time-reversal invariant BIs always transform according to the trivial irreducible representation. 
%
%{\color{C2}\textit{1.2 Proof distinct BI fMI.}} 
%The transformation behaviour of BIs can be derived using that for time-reversal invariant BIs Kramer's pairs are either fully occupied or unoccupied. Each Kramer's pair is formed by two single particle states whose symmetry eigenvalues multiply to one yielding an eigenvalue of one for the many-body state. Thus, if one finds a unique ground state that transforms non-trivially under some symmetry, it cannot be a BI, but is called (fragile) Mott insulator.
%
%
%
%{\color{C2}\textit{2.1 Hamiltonian.}}
We consider the four-site Hubbard ring at half filling described by the Hamiltonian
\begin{equation}\begin{split}
H =& - \sum_{j,\sigma} \left( t c_{j,\sigma}^\dagger c_{j+1, \sigma} + t' c_{j,\sigma}^\dagger c_{j+2 ,\sigma} + \mathrm{h.c.}\right) \\
& + U \sum_{j} c_{j, \uparrow}^\dagger c_{j, \uparrow} c_{j, \downarrow}^\dagger c_{j, \downarrow}~,\qquad \label{eq:ham}
 \end{split}
\end{equation}
where $c_{j,\sigma}^\dagger$ creates a fermion at site $j=1,2,3,4$ with spin $\sigma=\uparrow\!/\!\downarrow$, $t$ and $t'$ parametrize the nearest- and next-nearest-neighbour hopping, respectively, and $U$ is the on-site Hubbard interaction~\cite{fmi,fmi_titus}. The system is depicted in Fig.~\ref{fig:foursites}. 
%In dependence of the ratio $t'/t$ the ground state realizes either a BI or FMI, which can be distinguished using the point group symmetries.

%{\color{C2}\textit{2.2.2 Discuss Hamiltonian: Symmetries.}}
The spatial symmetry group of Hamiltonian~\eqref{eq:ham} is isomorphic to  $C_{4v}$ and generated by the four-fold rotation $C_4$ and mirror reflection $\mathcal{M}$ defined as
\begin{equation}
C_4 c_{j,\sigma} C_4^\dagger = c_{j+1,\sigma},\quad  \mathcal{M} c_{j,\sigma} \mathcal{M}^\dagger = c_{-j,\sigma}.
\end{equation}
For the transformation into the eigenbasis of $C_4$, the relation
%
%\begin{equation}
$c_{j,\sigma} = \frac{1}{2}\sum_{\lambda} \lambda^j \tilde{c}_{\lambda,\sigma}$
%\end{equation}
%
is used, where $\lambda$ runs over $\{\pm1,\pm i\}$. In addition, the Hamiltonian has time-reversal symmetry. Starting from the single-particle spectrum shown in Fig.~\ref{fig:foursites} for $U=0$, we now discuss the two cases $t' t>t/2$ and $t'<t/2$ for  small $U$.

%{\color{C2}\textit{2.2.1.2 Ground State Degeneracy Cases}}
For large next-nearest-neighbour hopping $t'/t > t/2$, the $U=0$ ground state is \emph{non-degenerate}, consisting of two occupied Kramer's pairs.
Due to the spectral gap, the ground state at small $U$ is adiabatically connected to a single Slater determinant. 
Conversely, for small next-nearest-neighbour hopping $t't < t/2$, the ground state at half filling is \emph{degenerate} at $U=0$. This degeneracy is lifted by finite $U>0$ and the unique ground state $(\tilde{c}^\dagger_{i,\uparrow}\tilde{c}^\dagger_{-1,\uparrow}\tilde{c}^\dagger_{i,\downarrow}\tilde{c}^\dagger_{-1,\downarrow}-\tilde{c}^\dagger_{-1,\uparrow}\tilde{c}^\dagger_{-i,\uparrow}\tilde{c}^\dagger_{-1,\downarrow}\tilde{c}^\dagger_{-i,\downarrow})|0\rangle/\sqrt{2}$ emerges to lowest order in $U/t$. The qualitative difference between the two regimes is evident from the symmetry eigenvalues of the respective ground states.

%{\color{C2}\textit{2.2.3 Application to Ground States}}
For $t'/t > 1/2$ the ground state has eigenvalues $\lambda=s_\mathcal{M}=+1$ for both $C_4$ and $\mathcal{M}$ symmetries, 
%respectively,  -ivano: why respectively if the value is the same?
thus belonging to the $A_1$ irreducible representation of $C_{4v}$. 
For $t'/t<1/2$ (and $U>0$) we have $\lambda=s_\mathcal{M}=-1$, placing the ground state in the $B_1$ irreducible representation~\cite{fmi}.
It can be shown that for a time-reversal-invariant spinful fermion system, a non-degenerate single--Slater determinant---or single-reference---ground state (and hence also any state adiabatically connected to one) has to be in the trivial irreducible representation of the spatial symmetry group. %The reason is that the single Slater determinant is composed of Kramers pairs which have complex conjugate eigenvalues under all spatial symmetries, thus multiplying to 1. 
The interest in the model given by Eq.~\eqref{eq:ham} is thus that for $t'/t<1/2$, $U>0$ its ground state is qualitatively different from any possible noninteracting state with the same symmetries. Due to this property, it can be used as a building block for two-dimensional fragile Mott insulators~\cite{fmi}, an intrinsically interacting quantum phase. %that represents a symmetry protected topological phase. 

\textit{Mapping to a quantum circuit.} 
%{\color{C2}\textit{3.1 Reasoning.}}
Our objective is to obtain and characterize this ground state transition on a current quantum computer in a scalable manner, meaning that the proposed procedure only involves algorithms scaling polynomially with the size of the system. %In principle, such an approach allows for an application to system sizes on future quantum computers beyond reach of classical computers. In this work, we focus on the four-site ring as a demonstration on current quantum hardware.

%{\color{C2}\textit{3.1 Choice of Trafo.}}
First, the fermionic Fock space needs to be mapped to a bosonic Hilbert space built of two-level systems, namely the qubits, which form the computational basis of the quantum computer. 
%Common mappings are the Jordan-Wigner (JW), parity, and Bravyi-Kitaev transformations~\cite{intro_qc,jw,parity,bk_map,bk_map_ext}. For these three options, the number of terms required for encoding the Hamiltonian given in Eq.~\eqref{eq:ham} is comparable.
 We choose the JW transformation since it provides a direct relation between the occupation of the fermionic states $n_{\lambda,\sigma}$ and qubit states $q_{k(\lambda,\sigma)} = n_{\lambda,\sigma}\in\{0,1\}$ with a chosen ordering $k:(\lambda,\sigma)\mapsto \{0,1,\ldots,7\}$. (Four sites with spin degeneracy amounts to an $2^{8}$-dimensional Hilbert space, which requires eight qubits, see Appendix~\ref{app:jordan_wigner}.)

%{\color{C2}\textit{3.2 Improve Jordan-Wigner Trafo:Tapering.}}
%{\color{C2}\textit{3.2.1 General Procedure.}}
The number of required qubits can be  reduced by means of tapering~\cite{taper,taper_molecule}. Tapering describes an efficient method of finding and exploiting a set of (mutually commuting) 
$\mathbb{Z}_2$ symmetries $\{\mathcal{S}\}$ of the Hamiltonian. The single-particle states  are chosen to form an eigenbasis of all $\mathcal{S}$ simultaneously in a way that  for each $\mathcal{S}$ there is a qubit $k_\mathcal{S}$ on which only $\mathcal{S}$ acts non-trivially. In particular, one chooses the basis such that the symmetry eigenvalues $s_\mathcal{S}\in\{\pm 1\}$ of the  many-body state are represented by the qubits $k_\mathcal{S}$ in the computational basis. 
In case the symmetries are preserved within the calculations, i.e., the calculation is restricted to one symmetry subspace, the eigenvalues $s_\mathcal{S}$ are fixed and can be used to replace the variables $q_{k_\mathcal{S}}$ in all calculations.

%{\color{C2}\textit{3.2.1 N=4 case}}
The Hamiltonian~\eqref{eq:ham} has four commuting $\mathds{Z}_2$ symmetries $\mathcal{S}\in \{C_2,\mathcal{M},\mathcal{P}_{\! \uparrow},\mathcal{P}_{\! \downarrow} \}$: the rotation $C_{2}=C_4^2$, the spatial mirror symmetry $\mathcal{M}$, and the parities of the number of up and down spins $\mathcal{P}_{\! \sigma}$. 
A common eigenbasis of these four symmetries  can be chosen such that four qubits $q_{k_\mathcal{S}}$ represent the eigenvalues of these symmetries (see Appendix~\ref{app:trafo_to_symm_eigenbasis}). Hence, they can be excluded from calculations on the quantum computer. This reduces the number of qubits required to represent the system from eight to four.
Note that both the JW mapping and tapering do not entail any overhead that scales exponentially with system size.

%{\color{C2}\textit{3.3 Relevant symmetry subspaces.}}
%For the purpose of showing the ground state transition on a current quantum computer, the symmetry eigenvalues $s_\mathcal{S}$ need to be selected to chose the corresponding symmetry subspaces. 
We study the model at half-filling and vanishing total spin. The symmetry eigenvalues separate the Hilbert space into sectors that correspond to states in the irreducible representations $A_1$, $B_1$, or $E$ of the point group $C_{4v}$. We will variationally compute the lowest-energy state in each of these sectors separately. (Note that for the $A_1$, $B_1$ irreducible representation, the discerning $C_4$ eigenvalue 
cannot be recovered after applying tapering.
Instead, we measure it using the relation $C_4=\prod_{\lambda,\sigma} \left( \lambda\right)^{n_{\lambda,\sigma}}$, for which the reverse transformation to tapering is required, see Appendix~\ref{app:undo_tap}.)
%We study the model at half-filling and vanishing total spin, yielding the four possible symmetry eigenvalues $s_{\mathcal{A}} = s_{\mathcal{B}}s_{\mathcal{M}}$, $s_{\mathcal{B}} = \pm 1$, $s_{\mathcal{M}}=\pm 1$ and $s_{\mathcal{P}_{\! \downarrow}}=1$. Of these, the two subspaces $s_{\mathcal{B}}=-s_{\mathcal{M}}=\pm1$ have the same spectrum: The only irreducible representation which allows for $s_\mathcal{B} s_\mathcal{M}=-1$ is the two-dimensional $E$, which implies a two-fold ground state degeneracy. 
%By choice of the basis the two degenerate states need to be eigenstates of $C_2$ and $\mathcal{M}$ with one state having $s_\mathcal{B}=-s_\mathcal{M}=1$ and one state $s_\mathcal{B} = -s_\mathcal{M}=-1$ such that $\chi(\mathcal{M}) = 0$ and $\chi(C_2) = -2$.

% %{\color{C2}\textit{3.3 Identifying the irrep.}}
% %In summary, we consider the three symmetry subspaces $(s_{\mathcal{B}},s_{\mathcal{M}})= (1,1),(-1,-1),(-1,1)$. 
% In the first two cases, the $C_4$ eigenvalue 
% %is not fully determined by tapering.
% cannot be recovered after applying tapering.
% To measure this value by means of the relation $C_4=\prod_{\lambda,\sigma} \left( \lambda\right)^{n_{\lambda,\sigma}}$, the reverse transformation to tapering is required (see SI~\ref{app:undo_tap}). Then, together with $s_\mathcal{M}$ and $s_\mathcal{B}$, the irreducible representation of the ground state can be deduced.

%{\color{C2}\textit{3.5.1 VQE}}
\textit{Variational algorithm.} 
The ground state is selected by comparing the smallest energy eigenstates of the system Hamiltonian in the different symmetry subspaces. In each subspace, the energies are estimated using the Variational Quantum Eigensolver (VQE)~\cite{vqe,vqe_google,general_vqe2}.
%, which relies on the variational principle. 
%This algorithm relies on the fact that the expectation value of the Hamiltonian with respect to any state $\Psi$ is bounded from below by the ground state energy $E_0$: $E_0 \leq \langle \Psi |H|\Psi\rangle$. 
% These concepts are too basics for PRL
Standard ans\"atze for the representation of the wavefunction range from adoptions of the variational forms used in quantum chemistry~\cite{UCCSD,vqe_google} to hardware efficient heuristic approaches~\cite{efficent_vqe,cxvscz}. 
An example for the latter is the $R_yR_z$ variational form, which consists of layers of $R_y$ and $R_z$ single qubit rotations on all qubits alternating with layers that entangle all qubits. 
Recently, adaptive circuits, which have entangling gates only between selected qubits, have been shown to be efficient %~\cite{evol_alg,adapt_vqe,evol_alg2} 
in reducing the total number of entanglement operations and hence the overall circuit noise~\cite{evol_alg,adapt_vqe,evol_alg2}. 

\begin{figure}[t]
\centering
\includegraphics{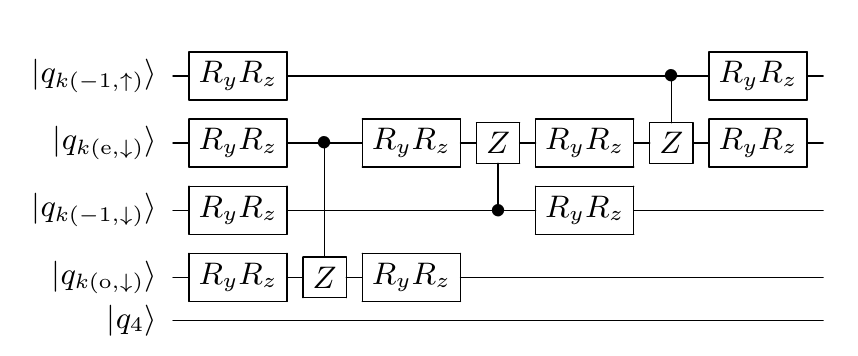}
\caption{Example of a sequential adaptive $R_yR_z$ circuit with $n_\mathrm{CZ}=3$ controlled $Z$ gates (CZ) adapted to the five-qubit backend \textit{ibmq\_ourense}. Solving the four-sited Hubbard ring requires only four qubits, hence, the eight initial rotations $R_y$, $R_z$ only act on the first four qubits. Three CZ gates are chosen from the set of all natively implemented controlled gates on the device. Each is followed by $R_yR_z$ rotations on the related qubits yielding in total $20$ variational parameters $\bm \Theta$, one angle $\Theta$ for each rotation $R_y$ or $R_z$.}
\label{fig:var_form}
\end{figure}

In this work, we explore the use of an adaptive $R_yR_z$ ansatz to describe the ground state of the system (see Fig.~\ref{fig:var_form}).
%motivating us to propose an adaptive $R_yR_z$ ansatz in order to reduce the number of entangling gates and, hence, the error. 
%
An initial layer of $R_yR_z$ rotations on all qubits is followed by a sequence of $n_\mathrm{CZ}$ entangling gates, namely controlled $Z$ gates (CZ). 
Each CZ is followed by $R_yR_z$ rotations parametrized with angles $\bm \Theta$ at the corresponding target and control qubits. 
%The sequence of rotations and CZs can be improved with an optimization routine for the angles $\bm \Theta$ as well as choosing the target and control qubits of the CZs.
%, we propose to adapt them to problem and the quantum computer architecture forming the adaptive $R_yR_z$ ansatz motivated by the $R_yR_z$.
%{\color{C2}\textit{3.6.2 Adaptive RyRz}}
%However, all of these require too many entangling (two-qubit) gates to obtain meaningful results on current quantum computer. 
%Here, we propose the use of an adaptive $R_yR_z$ ansatz, see also. It reduces the number of entangling gates further by adapting them to the . %
%The entangling gates, in this work controlled $Z$ gates (CZ),
%
%The target and control qubits of each of the CZs are 
The position of these qubits in the qubit register is chosen randomly among all natively implementable CZs in a given quantum  architecture, but then held fixed during $\boldsymbol \Theta$ optimization. 
%\textcolor{green}{you did not introduce the definition of $n_\mathrm{CZ}$ and the meaning of this sentence remains unclear to me: To obtain the best ground state estimate $n_\mathrm{CZ}$ and the number of variational parameters is adapted to the hardware capabilities.}

The ground state is obtained through variation of the single qubit parameters $\boldsymbol \Theta$ as well as by the simultaneous optimization of the CZ gates, i.e., their number $n_\mathrm{CZ}$ and position in the circuit.
%For fix $n_\mathrm{CZ}$, we therefore denote with $|\Psi^c(\bm \Theta)\rangle$ a generic trial wavefunction, where the parameter $c$ is labeling the $n_c$ realizations of the CZ network sampled with a stochastic optimization method, namely random search\cite{random_search}, until convergence (see SI~\ref{app:czseq}). 
%
For fixed $n_\mathrm{CZ}$, we therefore denote with $|\Psi^c(\bm \Theta)\rangle$ a generic trial wavefunction, where the index $c$ labels the different CZ configurations. The latter are optimized through a stochastic optimization method, namely random search \cite{random_search}, in a pool of $n_c$ possible circuits (see Appendix~\ref{app:czseq}).
%, which are compatible with a given hardware architecture.
%In summary, each variational form is fully characterised by a sequence $c$ of random CZ and the set of single qubit rotations $\boldsymbol \Theta$. A generic trail wavefunction is therefore denoted as $|\Psi^c(\bm \Theta)\rangle$. 
By minimizing the cost function
%$ E(\bm \Theta) =\langle \Psi(\bm \Theta) | H | \Psi(\bm \Theta) \rangle$ 
$L^c(\bm \Theta) = \langle \Psi^c(\bm \Theta) | H  + f (N-4)^2| \Psi^c(\bm \Theta)\rangle$ for a given ansatz $|\Psi^c(\bm \Theta)\rangle$, the optimal parameters $\bm \Theta_{\mathrm{opt}}^c$ and an upper bound for the ground state energy $E_{\mathrm{opt}}^c =  \langle \Psi^c(\bm \Theta_{\mathrm{opt}}^c) | H | \Psi^c(\bm \Theta_{\mathrm{opt}}^c) \rangle$ are obtained. 
The fully optimized wavefunction is then obtained by further minimizing in the space of all possible circuits at fixed $n_\mathrm{CZ}$, i.e., adapting the CZs with random search, which we denote by $\mathrm{min}_c$ below.
Note that the last term in the definition of $L^c(\bm \Theta)$ enforces the half filling condition, where $N$ is the number operator and $f$ is a hyperparameter which we fix to $f=0.05$.

The optimization of $\bm \Theta$  is performed classically using the COBYLA and SPSA~\cite{cobyla,spsa} algorithms.
%}
The initial values are sampled from an uniform distribution between $0$ and $2 \pi$. 
%The initial parameters are chosen as being the best out of $n_\mathrm{init}$ parameter sets, which are sampled from an uniform distribution between $0$ and $2 \pi$. 
For a given state $|\Psi_c(\bm \Theta)\rangle$, % with CZ sequence $c$ parametrized by $\bm \Theta$, % $L_\mathrm{seq}(c,\bm \Theta)$ 
the corresponding expectation value of the system Hamiltonian and the number-operator term can be evaluated efficiently on a quantum computer. 
The optimization (i.e., classical update of the coefficients and measurement of the energy) is carried on until convergence (see Appendix~\ref{app:cobyla_spsa}).
%The quantum computer results are used by the optimization method to repeatedly suggest new parameters for which $L_\mathrm{seq}$ is evaluated again until convergence. 
%The combination of a quantum computer and a classical computer in each iteration yields an algorithm which does not rely on a vector formulation in the exponentially large Hilbert space. 
%Thus, it is expected to be applicable efficiently to larger systems; the exact scaling of the method can be estimated only with sufficiently large quantum computers and depends sensitively on the choice of variational form.
%
%
%In this work, COBYLA is used for \emph{non-noisy} simulations (no-n), in which expectation values are evaluated 
%
%with the vector representation for states and the matrix representation for operators and gates. 
%
%within the statevector or matrix representation as implemented in Qiskit \cite{qiskit}. On the other hand, we use SPSA for stochastic simulations and calculations on quantum computers, due to its better performance in these situations~\cite{cobylagood,spsagood,efficent_vqe}.
%
%Recent studies found the two methods to perform well for these respective usecases~\cite{cobylagood,spsagood,efficent_vqe}.
%
%{\color{red}
%In a subsequent step the sequence of random CZ $c$ is improved by performing the VQE optimization routine for each of $n_c$ ans\"atze $\{|\Psi^c(\bm \Theta)\rangle\}_{c=1}^{n_c}$ with fixed $n_\mathrm{CZ}$. We then choose the sequence $c$ with the best performance as measured by the objective function $L^c(\bm \Theta^c_\mathrm{opt})$, indicated with $\mathrm{min}_c$}.
%
\begin{figure*}[t] 
\includegraphics[scale=1]{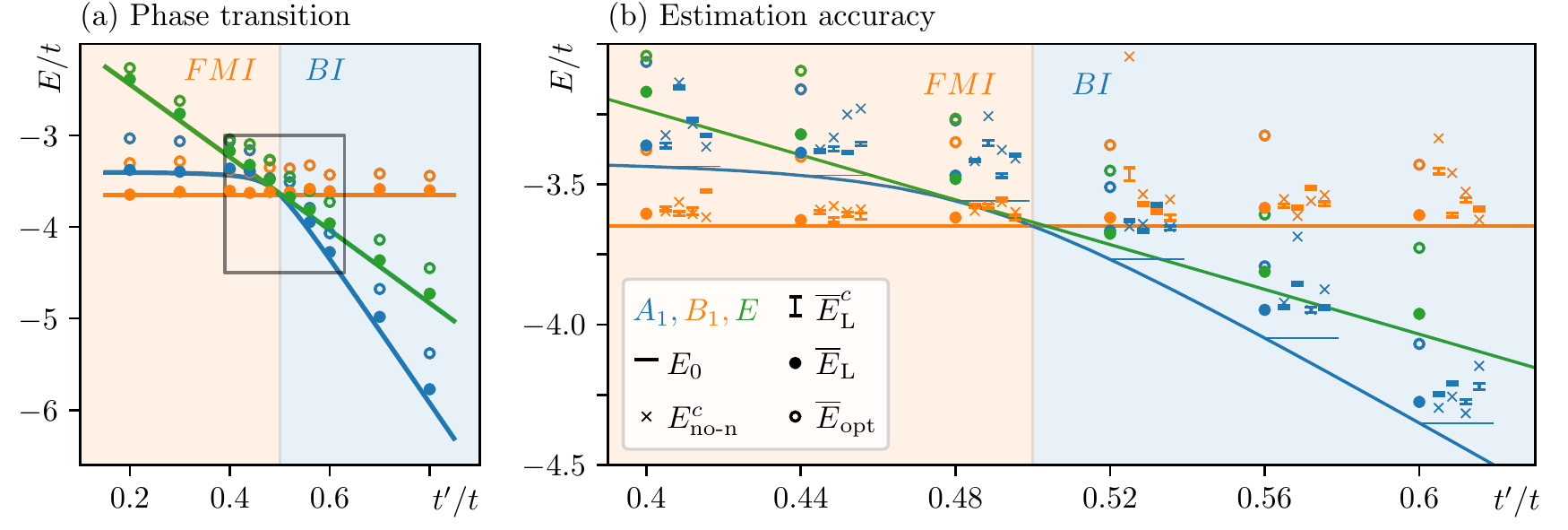}
\caption{\label{fig:phase_transition_with_inset}
The results for the ground-state energy estimates at $U=t/2$ are shown for the three symmetry subspaces: in orange for $B_1$, in blue for $A_1$ and in green for $E$. The lines denote the exact diagonalisation ground-state energy. The error bars visualise the weighted least square Lanczos result for each CZ sequence (see Appendix~\ref{app:czseq}), with $n_\mathrm{CZ}=3$. The crosses mark the ground-state estimate $E_{\text{no-n}}^c$ using the non-noisy simulation for the state optimized on the quantum computer. The dots represent the minimum-weighted least-square Lanczos result of the four CZ sequences. For comparison, the minimum-weighted least-square result of $E_{\mathrm{opt}}^c$ is shown with circles with error bars too small to be shown. For better readability, the data for each of the $n_c=4$ sequences is shown right to the minimum result they belong to.}
\end{figure*}
%
%\textit{Statistical error mitigation.} 
%weighted least squares

\textit{Quantum error mitigation.} 
%{\color{C2}\textit{3.9 Lanczos}}
Using the optimization procedure, we obtain variational estimates $E_{\mathrm{opt}}^c$ %$\langle H \rangle_{\mathrm{qc}}$ 
for the ground-state energies% for each point $U/t,t'/t$
, which are compromised by noise and errors in the quantum device. 
To counter-act this problem, various error mitigation schemes have been proposed~\cite{qiskit,error1,error2,error3,error4,Note2}\footnotetext[2]{IBM Research, “Qiskit Ignis,” https://qiskit.org/ignis
(2019).}. 
In this work, we use a Lanczos-inspired error mitigation algorithm as it does not yield any additional systematic errors, does not scale exponentially with system size, and has been shown to improve the accuracy of the results significantly~\cite{suchsland2020algorithmic}. 

The measurement uncertainty of the mitigated results $E_{\mathrm{L}}^c$ and unmitigated results $E_{\mathrm{opt}}^c$ is reduced by calculating $K$ estimates $E_{\mathrm{L},l}^c$ or $E_{\mathrm{opt},l}^c$ with $l\in\{1,2,\ldots,K\}$
and measurement uncertainties $\sigma_l$. These are averaged with weighted least squares
\begin{equation}
\overline{x} =  \left. \sum_{l=1}^{K} x_l  \sigma_l^{-2} \middle/  \sum_{l=1}^K  \sigma_l^{-2}\right. ,% \qquad %\sigma_{\mathrm{\omega}[x]}^{2} =1\Big/ \sum_l \sigma_l^{-2}
\end{equation}
yielding estimates for the ground-state energy. In the following, we use $K=5$.

\textit{Results.} 
%{\color{C2}\textit{4.1.2 Simulation results}}
In order to validate the method and to optimize hyperparameters, we first perform simulations on a classical computer without noise.
%with a realistic simulation of the hardware noise. 
%\textcolor{green}{REMOVE - this sentence is strange. What is the upper bound you are talking about? To justify the use of the upper bounds as ground state estimates, we perform non-noisy simulations. As a condition, we require an upper bound close to the exact ground state energy in comparison to the energy of the first excited state. }
The quality of the ground-state energy estimates is assessed by comparing the error with the energy gap to the first excited state.
We analyse the accuracy of the ground state VQE energies for $U=t/2$ and $t'/t \in [0,1]$ comparing the variational forms q-UCCSD, $R_yR_z$ and the adaptive $R_yR_z$. 
For all variational forms, a ground-state energy accuracy at least one order of magnitude smaller than the excitation gap to the first excited state can be obtained. 
%\textcolor{green}{[REMOVE] What is the meaning of this sentence? For up to $n_\mathrm{CZ}=15$ entangling gates, the adaptive $R_yR_z$ circuit yields the best results: }
For the adaptive $R_yR_z$ the accuracy is the highest and improves with increasing circuit depth $n_\mathrm{CZ}$ up to $\langle |E_{\mathrm{opt}}-E_0|/t \rangle \approx 10^{-4}$ for $n_\mathrm{CZ}=15$, 
where the brackets $\langle \dots \rangle$ indicate an average over the two symmetry subspaces relevant for the ground state transition and over values $t'/t$ $\in [0, 1]$ (see Appendix~\ref{app:comp_var_forms}).
%(see SI~\ref{app:comp_var_forms}).
%}
%\textcolor{green}{What is averaged? averaged over the two symmetry subspaces relevant for the ground state transition and over $t'/t$ ranging from $0$ to $1$.}

%{\color{C2}\textit{4.2 Discussion of the results}}
%{\color{C2}\textit{4.2.1 Phase Transition}}
The results for $\overline{E}_\mathrm{opt}=\mathrm{min}_c\overline{E}^c_\mathrm{opt}$ obtained with the IBM quantum computer \textit{ibmq\_ourense} % and the enhancements with the Lanczos algorithm 
are shown in Fig.~\ref{fig:phase_transition_with_inset} \cite{Note3}\footnotetext[3]{In stochastic simulations~\cite{qasm} with a realistic simulation of the hardware noise we obtain no significant performance improvement by going beyond an adaptive $R_yR_z$ variational circuit with $n_\mathrm{CZ}=3$, a limit of $100$ optimization steps with $1024$ shots for the expectation value estimates, $n_\mathrm{init}=5$ and $n_c=4$. Hence, these settings are used for calculations on the quantum devices.}. 
The left panel shows that the ground state transition is correctly predicted for $t'/t$ values between 0.48 and 0.52.
%\textcolor{green}{all this is very confusing since \omega was introduced in Eq 4 in relation to Lanczos. Here you use it with min$_c$ which is not well explained. I will postpone the introduction of Lanczos until the paragraph starting with "In order to reduce ..." }

%\textcolor{green}{I stop commenting at this point.}

%, where we take the minimum of the result for the $n_c=4$ sequences.
In general, three different sources of error can be analysed with the results in Fig.~\ref{fig:phase_transition_with_inset} (b). (\textit{i}) The measurement error induces an uncertainty in the evaluation of $\overline{E}_{\mathrm{opt}}^c$, 
which amounts to an average of $5\cdot 10^{-3} \, t$ and is the 
smallest among the three considered. 
(\textit{ii}) The variation of the results due to the stochastic optimisation routine 
%for $\mathrm{min}_c\, \mathrm{\omega}[\langle H \rangle_{\mathrm{qc}}]$
is estimated to be one order of magnitude larger using the $t'/t$ independent results of the $B_1$ subspace at different $t'/t$. (\textit{iii}) The systematic error, i.e., the offset to the true ground state energy, averages to about $0.4\, t$ thereby dominating over (\textit{i}) and (\textit{ii}).

To understand the source of (\textit{iii}), we consider the energy expectation value obtained in absence of  noise, $E_{\text{no-n}}^c$, using the optimal parameter set $\bm \Theta_{\mathrm{opt}}^c$ from the quantum computation [the same set used for Fig.~\ref{fig:phase_transition_with_inset} (b)]. 
In general, the $E_{\text{no-n}}^c$ are notably closer to $E_0$ than the result evaluated as $\overline{E}_{\mathrm{opt}}^c$. 
We therefore conclude that the dominant contribution to (\textit{iii}) the systematic error is hardware noise (e.g. gate errors, thermalization errors, readout errors)~%(\textcolor{green}{too vague, which other noise sources are possible? which one of the QC noise are you referring to?}) 
\cite{Note4}\footnotetext[4]{At $t'/t\approx 1/2$ for $A_1$ the systematic error due to the noise is less prominent since an avoided level crossing occurs, making the approximation of the ground state more difficult and resulting in a relatively smaller approximation accuracy of the VQE.}.
%
%We note, that the approximation of the ground state for $A_1$ and large $t'/t$ is smaller than expected as the ground state dominantly consists of a product state. This can be explained with the optimization routine which did not converge in this regime within $100$ optimization steps and the large gap to the first excited state. A small deviation from the true ground states yields worse energy estimates compared to a situation with a narrow spectrum.
%
The systematic error is found to be larger than the energy gap between the different states involved in the transition, as shown in Fig.~\ref{fig:phase_transition_with_inset}, so that a reliable prediction of the transition is only possible if the systematic error remains constant over the sampled parameter space.
%$relies on a constant systematic error for all results. 

%{\color{C2}\textit{4.2.2 Phase Transition Lanczos}}
In order to reduce the impact of the hardware noise and to overcome the dependency on the approximately constant systematic error, we use the Lanczos algorithm (see Fig.~\ref{fig:phase_transition_with_inset}). 
%In general, for these results we observe that the performance of the four different entangling sequences might vary considerably, although many entangling sequences yield similarly good results. Hence, for this case a smaller $n_\mathrm{config}$ could have been used, while in other cases an optimization of the entangling gates is expected to be crucial.
%The lowest of the four results obtained with different  entangling sequences $\mathrm{min_{seq.}\,\omega}[E_\mathrm{qc}]$ is shown in Fig.~\ref{fig:phase_transition_with_inset} (a). 
In comparison to 
%the non-noisy (\textcolor{green}{no-noise?}) variational results
$E_{\text{no-n}}^c$, the estimates obtained with the Lanczos algorithm, $\overline{E}_{\mathrm{L}}^c$, are of a similar or better accuracy considering the three already discussed sources of error. (\textit{i}) The measurement uncertainties of  $\overline{E}_{\mathrm{opt}}^c$ and  $\overline{E}_{\mathrm{L}}^c$ are similar. (\textit{ii}) The variation and (\textit{iii}) the systematic error are reduced by a factor three through the Lanczos algorithm. %systematic error has  factor three to four. 
As a result, the energy differences between the estimated values $\overline{E}_{\mathrm{L}}=\mathrm{min}_c\overline{E}_{\mathrm{L}}^c$  and the exact ground-state energy become smaller than the energy gaps,  
%for the exact ground state energy $E_0$ is substantially smaller than the energy gap between the levels
allowing for the accurate detection of the ground-state transition. 
Within the measurement uncertainty, the obtained ground state lies in the correct symmetry subspace over the entire parameter space (as confirmed by comparison to exact diagonalization). 
Our approach allows to resolve energy differences up to $0.1 \, t$, enabling the detection of ground-state symmetry breaking and transitions in a scalable manner on noisy near-term quantum computers. In classical simulations we obtained the ground state transition in the six-sited molecule with the proposed scheme which we therefore expect to work on a future quantum computer with less hardware noise (see Appendix~\ref{app:six_sites}).
%(see SI~\ref{app:six_sites}).

\section{Conclusion}
%\textit{Summary.---} 
We presented a workflow to  solve variationally for the ground state of many-body quantum systems with quantitative accuracy on current quantum computing hardware. 
In the case of a four-site fermionic Hubbard ring, we demonstrated that our approach allows to detect the transition in the character of the ground-state solution. These results are enabled by the combination of a suited symmetry reduction of the problem, the application of a hardware-efficient variational ansatz, and a use of a Lanczos-inspired error mitigation algorithm.
Our work constitutes a benchmark on the way to simulations of many-body quantum systems beyond the reach of classical computers.

%This way we have shown the usefullness of near term quantum computers for the characterization of unknown ground states. With progress being made this can path the way for the expansion of the current knowledge about non-integrable systems beyond the reach of classical computers.

%- given full scheme for obtaining and characterizing systems on quantum computers scaling
%- for the four sited molecule we have shown, that on a perfect quantum computer the scheme is able to describe the ground state.
%- on real quantum computer we adapt the scheme to noise and limited computing ressources when dealing with NISQ devices
%- allows for obtaining the ground state transistion for the four sided molecule successfully
%- might be the starting point for future simulations allowing to determine black box ground state systems properties

%Outlook:
%- simulate larger system if q.c. allows
%- use more advanced techniques for the generic algorithms
%- simulate other systems

\vspace {1cm}

\section{Appendix}

\subsection{Jordan-Wigner Transformation}\label{app:jordan_wigner}
%{\color{C2}\textit{3.1 Jordan-Wigner Trafo.}}
The occupation of each fermionic state $n_{\lambda,\sigma} \in\{0,1\}$ is mapped to the value of a qubit $q$ which also takes the two values $\{0,1\}$ corresponding to the two levels of the two-level system. Thus, a state $\Psi$ in the Fock space with $M=2N$ (for spin degeneracy) fermionic states is encoded in $M$ qubits
\begin{equation}
 |\Psi\rangle = |n_{M-1},n_{M-2},\dots,n_0\rangle \stackrel{\text{JW}}{\rightarrow} |q_{M-1},q_{M-2},\dots,q_0\rangle,
\end{equation}
via $q_{\lambda,\sigma} = n_{\lambda,\sigma}$ for each fermionic state and with a chosen bijection $(\lambda ,\sigma) \mapsto k \in\{0,1,\ldots,M-1\}$. To restore the right commutation relations, the fermionic creation and annihilation operators with canonical anti-commutation relations are mapped to the spin lowering and raising operators with canonical commutation relations using
\begin{eqnarray}
\tilde{c}_k \stackrel{\text{JW}}{\rightarrow}&c_{k} &= A_k Z_{k-1}\cdots Z_0,\nonumber\\
\tilde{c}^\dagger_k  \stackrel{\text{JW}}{\rightarrow}&  c_{k}^\dagger &= \left(c_{k}\right)^\dagger = {A_k}^\dagger Z_{k-1}\cdots Z_0,
\end{eqnarray}
with $A_k := \frac{1}{2} (X_k + iY_k )$ and $Z_k|q_k\rangle = (-1)^{q_k}|q_k\rangle$,  where the operations on the $k$th qubit are denoted by the Pauli matrices $X_k,Y_k,Z_k$ and the identity $\mathds{1}_k$.

\subsection{Unitary Transformation into the Symmetries Eigenbasis} \label{app:trafo_to_symm_eigenbasis}
An eigenbasis of all symmetries is formed by the fermionic states $d_{\mathrm{e},\sigma} = \left( \tilde{c}_{i,\sigma} + \tilde{c}_{-i,\sigma}\right)/\sqrt{2}$, $d_{\mathrm{o},\sigma} = \left( \tilde{c}_{i,\sigma} - \tilde{c}_{-i,\sigma}\right)/\sqrt{2}$ and $d_{\pm1,\sigma} = \tilde{c}_{\pm1,\sigma}$. Applying the JW transformation to $d_{\mathrm{e/o},\sigma},d_{\pm1,\sigma}$, the four symmetries are representable as a specific tensor product of $\mathds{1}_k$ and $Z_k$ each. With $\mathds{1}_k,X_k,Y_k,Z_k$ we refer to the  identity and the respective Pauli operators acting on qubit $k$. Recombining the symmetries to $\mathcal{A} = C_2 \mathcal{P}_{\! \uparrow}$, $\mathcal{B} = C_2 \mathcal{M}$, $\mathcal{M}$ and $\mathcal{P}_{\! \downarrow}$ allows to choose four qubits $k_\mathcal{A},k_\mathcal{B},k_\mathcal{M}$ and $k_{\mathcal{P}_{\! \downarrow}}$, on which only the corresponding symmetry acts non-trivially. Finally, the unitary transformation $U=U_{\mathcal{P}_{\! \downarrow}} U_{\mathcal{M}} U_\mathcal{B} U_\mathcal{A}$ with $U_\mathcal{S}=(X_{k_\mathcal{S}} + \mathcal{S})/\sqrt{2}$ is used to transform into the basis where the $q_{k_\mathcal{S}}$ store the values $s_\mathcal{S}$. Hence, they can be excluded from calculations on the quantum computer. This reduces the number of qubits required to represent the system from eight to four.

\subsection{Undo Tapering Procedure and Measurement of the Symmetry Eigenvalue $\lambda$}\label{app:undo_tap}
\paragraph{Undo Tapering Procedure}
To undo tapering for a symmetry $\mathcal{S}$, first, a qubit is added for the tapered qubit $k_\mathcal{S}$ in state $|(1-s_\mathcal{S})/2\rangle$. The next step relies on the observation that all symmetries are representable as one tensor product of $Z$ Pauli matrices and identities~\cite{taper}. For all $Z_k$ in the tensor product a CNOT with qubit $k$ as a control and $k_\mathcal{S}$ as a target is applied. This yields the measurement outcome of $s_\mathcal{S}$ for the expectation value of $\mathcal{S}$. It ensures that $q_{k_\mathcal{S}}$ encodes the same value compared to before the tapering procedure since it is determined by the $\mathbb{Z}_2$ symmetry. In a final step the transformation in the eigenbasis of the mirror symmetry can be reversed with a sequence of two-qubit unitaries $V$, one unitary for each pair of states connected through the mirror symmetry. Note, that the transformation depends on the chosen order $k$ for the Jordan-Wigner transformation to obtain the right fermionic ordering. Hence, an ordering is favourable where the states forming a pair are close to each other. In the case of the four-site molecule $V$ transforms $d_{e/o,\sigma}$ back to $\tilde{c}_{\pm i,\sigma}$. In this basis, the quasi-momentum is calculated by measuring $q_{k(\lambda,\sigma)}$:
\begin{equation}
    C_4=\frac{1}{4}\prod_\sigma Z_{k(-1,\sigma)} (1-iZ_{k(i,\sigma)}) )(1+iZ_{k(-i,\sigma)} ).
\end{equation}
%The values of the still tapered qubits $k_\mathcal{S}$ is determined by the symmetries along with the fixed symmetry eigenvalues. 

\begin{figure}[t]
\centering
\includegraphics[scale=1]{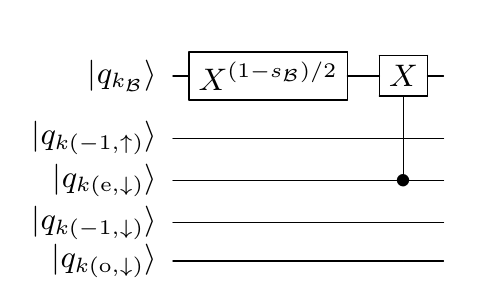}
\caption{The procedure to undo the tapering method is shown in this figure for the symmetry $\bm \sigma_\mathcal{B}$ as an example. First, an ancilla qubit $q_{k_\mathcal{B}}$ in the state $|(1-s_\mathcal{B})/2\rangle$ is added to the qubits $q_0-q_3$ which were used for the calculations before. Then, for each $Z_k$ gate in the tensor product representation of the symmetry $\bm \sigma_\mathcal{B}=Z_{q_{k_\mathcal{B}}} Z_{k(e,\downarrow)}$, a controlled $X$ gate is applied on $q_{k_\mathcal{B}}$ with $q_k$ as control. In this case, there is only one $Z_k$ gate, namely $Z_{k(e,\downarrow)}$, yielding one controlled $X$ gate as shown in the figure.}
\label{fig:rev_taper}
\end{figure}

A mathematical formulation of the procedure to reverse the tapering is given in Eq.~\eqref{eq:rev_taper}. In this, we show that the state reconstructed as described in the main text and Fig.~\ref{fig:rev_taper} yields the desired symmetry eigenvalue $s_\mathcal{S}$. 

We denote with $\mathcal{S}_k$ the part of $\mathcal{S}$ acting on qubit $k$ based on the assumption that $\mathcal{S}$ has a tensor product structe. This was observed in this paper and the original reference~\cite{taper}. The added qubit in state $|0\rangle$ is given as $|0\rangle_{k_\mathcal{S}}$ with the undo tapering operator $R$. The evaluation of the symmetry operation $\mathcal{S}$ on the state after reversing the tapering procedure yields
%
%use widetext to get an equation over the whole page
\begin{widetext}
\begin{eqnarray}
\mathcal{S} R|q_{M-1},\dots,q_0\rangle\otimes|0\rangle_{k_\mathcal{S}} 
&=& \left(\prod_{k:\mathcal{S}_k=Z_k}(-1)^{q_k} \right) |q_{M-1},\dots,q_0\rangle \otimes Z_{k_\mathcal{S}}\left(X_{k_\mathcal{S}}\right)^{(1-s_{\mathcal{S}})/2}\left(\prod_{k:\mathcal{S}_k=Z_k}(X_{k_\mathcal{S}})^{q_k} \right)|0\rangle_{k_\mathcal{S}}\nonumber \\
&=& s_{\mathcal{S}}R | q_{M-1}, \dots, q_0\rangle\otimes|0\rangle_{k_\mathcal{S}}. \label{eq:rev_taper}
\end{eqnarray}
\end{widetext}
Hence, the state stores the correct symmetry eigenvalue for $\mathcal{S}$. Therefore, the system is in the same state as it has been fixed to by the tapering procedure.

\begin{figure}[t] 
\includegraphics[scale=1]{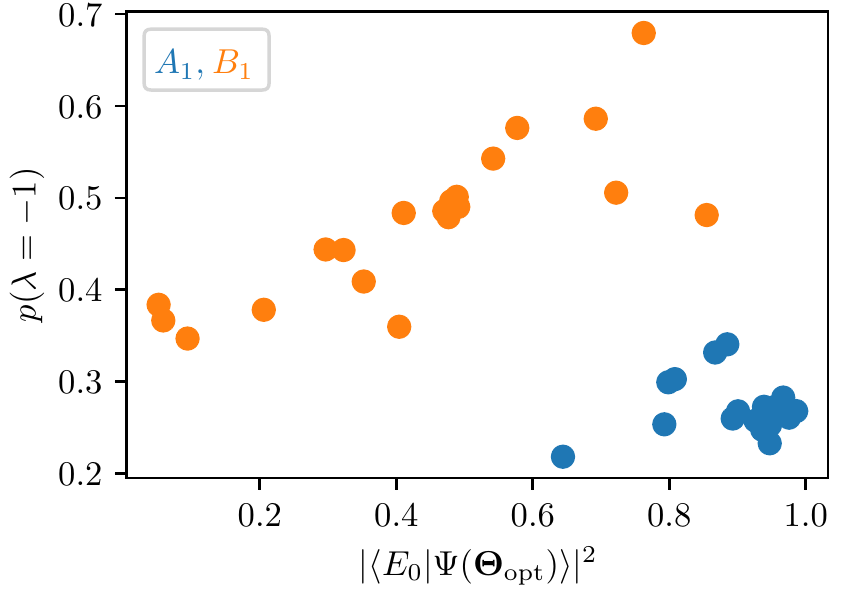}
\caption{\label{fig:symmetry_vs_overlap1}Result of the measurement of the momentum on the quantum computer \textit{ibmq\_ourense}. We show the probability of measuring $\lambda=-1$ for the ground state for all results shown in Fig.~\ref{fig:phase_transition_with_inset} in the symmetry subspaces with $s_{C_2}=1$, where $\lambda \in\{1,-1\}$. The value $|\langle E_0 | \Psi(\bm \Theta_\mathrm{opt})\rangle|^2$ denotes the overlap of the ground state with the learned state of the VQE in a non-noisy calculation. This shows a correlation of the measurement outcome with the overlap, as it approaches $1$ for $B_1$ and $0$ for $A_1$ for an overlap of $1$, as expected.}
\end{figure}

\paragraph{Measurement of the Symmetry Eigenvalue}
In Fig.~\ref{fig:symmetry_vs_overlap1} the results for the measurement of the symmetry eigenvalue $\lambda$ are shown. Errors occurring on the quantum computer are mitigated by disregarding unphysical results, i.e. measurements with the wrong already known symmetry eigenvalues. In simulations, this error mitigation schemes works at least as good as the Lanczos algorithm with the advantage of being computationally less demanding. The results are shown in Fig.~\ref{fig:symmetry_vs_overlap1}. 
%
%\begin{figure}[t] 
%\includegraphics[scale=1]{figures/momentum_tp_t.pdf}
%\caption{\label{fig:meas_symmetry}The result of the %measurement of the momentum is shown. We show the %fraction of measurements of $\lambda \in\{-1\}$ as in %the symmetry subspaces with $s_{C_2}=1$, $\lambda %\in\{1,-1\}$. }
%\end{figure}
%
%
%Here we can observe the following. For large $t'/t$ in the $A_1$ symmetry subspace, there is a clear indication of a total momentum of $0$. 

For $B_1$ the first excited state in the symmetry subspace is to close to the ground state to be able to distinguish them with the noise of the quantum computer. Effectively, these states are degenerate. Hence, the VQE method finds superpositions of these states. This can be seen in Fig.~\ref{fig:symmetry_vs_overlap1}, where the momentum measurement is shown in dependence of the overlap with the ground state. Here, we can observe the expected solutions in case the overlap with the ground state is large.

\subsection{CZ Sequences} \label{app:czseq}

For the optimisation of the CZ sequences stochastic optimization, namely random search \cite{random_search}, was used due to the limited quantum computing resources. We start with four initial random CZ configuration for each of which we perform the VQE optimization. Then we use the configuration out of the four with the smallest ground state energy estimate. For more complex applications this procedure might be extended to genetic algorithms with more generations and more local mutations, as also described in~\cite{evol_alg,adapt_vqe,evol_alg2}. Note, that all four generated wavefunction ans\"atze in all cases are found to describe states at half-filling. 

The randomly generated CZ sequences are described by the CZ seq. data in Tab.~\ref{tab:dataruns1}-\ref{tab:dataruns3}. It refers to the used CZ sequences in the main text for the variational forms $|\Psi^c(\boldsymbol{\Theta})\rangle$, we always used three CNOTs in a row. The numbers correspond to the control qubits and target qubits on the IBM quantum device \textit{ibmq\_ourense} as shown in Tab.~\ref{tab:ctrltargq}.
\begin{table}
    \centering
    \begin{tabular}{c|c|c}
         number & control q. & target q.  \\ \hline
         0 & 0 & 1 \\
         1 & 1 & 0 \\
         2 & 1 & 2 \\
         3 & 1 & 3 \\
         4 & 2 & 1 \\
         5 & 3 & 1 \\
    \end{tabular}
    \caption{The control and target qubits are denoted which are assigned to each number used in Tab.~\ref{tab:dataruns1} - \ref{tab:dataruns3}.}
    \label{tab:ctrltargq}
\end{table}

For example, the CNOT configuration $012$ corresponds to the first CNOT acting on qubit $1$ with qubit $0$ as control, the second CNOT acting on qubit $0$ with qubit $1$ as control and the last qubit acting on qubit $2$ with qubit $1$ as control.

\begin{table}
\centering
%\begin{minipage}{0.3\textwidth}
\begin{tabular}{l|llcccll}
 $t_2$  & CNOT seq. $c$ \\ \hline
 $0.2$  & 505 441 031 454 \\ % & 0  & 2/.239   \\ 
 $0.4$  & 300 513 010 335 \\ % & 1  & 4/.242   \\
 $0.6$  & 444 145 155 235 \\ % & 2  & B/.253   \\
 $0.8$  & 132 525 245 232 \\ % & 3  & C/.253   \\
 $0.44$ & 355 340 134 150 \\ % & 0  & 3/.240   \\
 $0.48$ & 443 432 425 305 \\ % & 1  & 4/.242   \\
 $0.52$ & 021 145 021 055 \\ % & 2  & E/.26    \\
 $0.56$ & 054 511 004 554 \\ % & 3  & C/.253   \\
 $0.3$  & 044 134 150 325 \\ % & 0  & 1/.240   \\
 $0.7$  & 151 332 511 155 \\ % & 1  & 1/.242   \\
\end{tabular}
\caption{Times for data for the antisymmetric subspace $B_1$.  The data was generated between $7^\mathrm{th}$ and $25^\mathrm{th}$ of March, 2020.}
\label{tab:dataruns1}
%\end{minipage}
\end{table}
%~
%\begin{minipage}{0.3\textwidth}
\begin{table}
\centering
\begin{tabular}{l|llcccll}
 $t_2$  & CNOT seq. $c$ \\ \hline
 $0.2$  & 105 301 452 032\\ % & 0  & 1/.238   \\ 
 $0.4$  & 430 441 313 344\\ % & 1  & 5/.243   \\
 $0.6$  & 111 353 020 254\\ % & 2  & 3/.240   \\
 $0.8$  & 222 323 310 451\\ % & 3  & A/.251   \\
 $0.44$ & 530 151 003 534\\ % & 0  & A/.251   \\
 $0.48$ & 532 200 445 251\\ % & 1  & 5/.243   \\
 $0.52$ & 435 244 203 400\\ % & 2  & B/.252   \\
 $0.56$ & 015 355 535 453\\ % & 3  & A/.251   \\
 $0.3$  & 045 045 451 434\\ % & 0  & 1/.238   \\
 $0.7$  & 330 102 124 153\\ % & 1  & 1/.239   \\
\end{tabular}
\caption{Times for data for the symmetric subspace $A_1$. The data was generated between $7^\mathrm{th}$ and $26^\mathrm{th}$ of March, 2020.}
\label{tab:dataruns2}
%\end{minipage}
\end{table}
%~
%\begin{minipage}{0.3\textwidth}
\begin{table}
\centering
\begin{tabular}{l|llcccll}
 $t_2$  & CNOT seq. $c$ \\\hline
 $ 0.3$ & 200 342 004 204 \\ % & 0  & 1/.238   \\
 $0.7$ & 225 215 333 415  \\ % & 1 & B/.252   \\ 
 $0.44$ & 120 304 041 345 \\ % & 0 & 3/.240   \\ 
 $0.48$ & 112 130 152 032 \\ % & 1 & A/.251   \\ 
 $0.52$ & 001 443 025 035 \\ % & 2 & 5/.243   \\ 
 $0.56$ & 323 344 241 512 \\ % & 3 & 1/.238   \\ 
  $0.2$ & 404 344 533 431 \\ % & 0 & 3/.240   \\ 
  $0.4$ & 341 350 452 401 \\ % & 1 & 5/.243   \\ 
  $0.6$ & 240 012 441 305 \\ % & 2 & A/.251   \\ 
  $0.8$ & 051 401 421 200 \\ % & 3 & B/.252   \\ 
\end{tabular}
\caption{Times for data for the odd momentum subspace $E$. The data was generated between $8^\mathrm{th}$ and $18^\mathrm{th}$ of April, 2020.}
\label{tab:dataruns3}
%\end{minipage}
\end{table}

\subsection{Classical Optimization Procedure} \label{app:cobyla_spsa}
The quantum computer results are used by the optimization algorithm to repeatedly suggest new parameters.
%for which $L_\mathrm{seq}$ is evaluated again until convergence. 
%The combination of a quantum computer and a classical computer in each iteration yields an algorithm which does not rely on a vector formulation in the exponentially large Hilbert space. 
%Thus, it is expected to be applicable efficiently to larger systems; the exact scaling of the method can be estimated only with sufficiently large quantum computers and depends sensitively on the choice of variational form.
%
%
In this work, the algorithm Constrained Optimization By Linear Approximations (COBYLA) is used for \emph{non-noisy} simulations (no-n), in which expectation values are evaluated within the statevector or matrix representation as implemented in Qiskit \cite{qiskit}. On the other hand, we use the algorithm Simultaneous Perturbation Stochastic Approximation (SPSA) for stochastic simulations and calculations on quantum computers, due to its better performance in these situations~\cite{cobylagood,spsagood,efficent_vqe}.

\subsection{Comparison of the Variational Forms}\label{app:comp_var_forms}
\begin{figure}[b]
\includegraphics[scale=1]{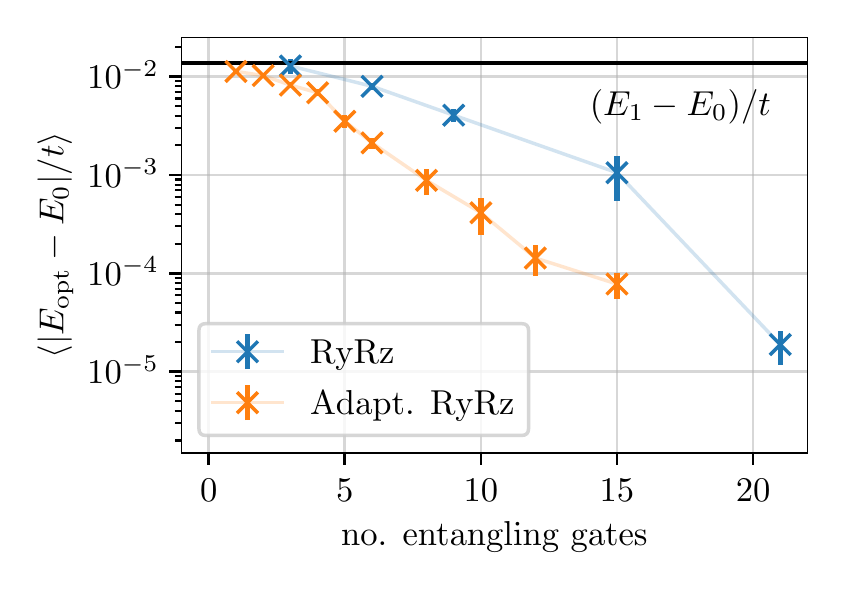}% Here is how to import EPS art
\caption{\label{fig:non-noisy}The results of the non-noisy simulations for different circuit depths are shown. The generation of the data is done as explained in the main text. To obtain stable ground state energy estimates for each considered $t'/t$ and symmetry subspaces $20$ VQE runs are performed and the lowest value is taken as a final result. These are averaged over $t'/t$ and the symmetry subspaces to obtain the data shown. With a vertical line the minimal excitation gap within the symmetry subspaces of all points $t'/t$ considered is shown.}
\end{figure}

In this section the results of the comparison of the different variational approaches are shown in Fig.~\ref{fig:non-noisy} for all investigated depths of entangling gates. The results indicate, that the adaptive variational form performs better for small numbers of entangling gates compared to the $R_yR_z$ variational form with linear entanglement. Furthermore, as stated in the main text, we can observe that the accuracy is high in comparison to the excitation gap from which we can deduce that the state generated by optimized variational form has a high overlap with the actual ground state of the system.

\subsection{Transition for the six-site Molecule} \label{app:six_sites}
\begin{figure}[t]
\includegraphics[scale=1]{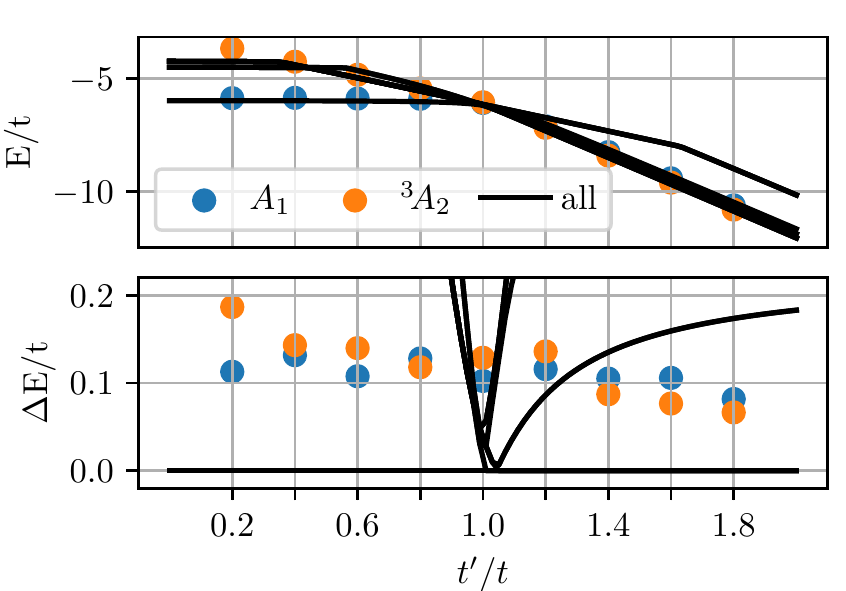}% Here is how to import EPS art
\caption{\label{fig:results_n6}
The results of the VQE optimization for the six-site molecule at $U=1.5t$ %,\phi=0$
are shown. The VQE optimization was performed for the two symmetry subspaces involved in the phase transition. In the top panel, the obtained energies are shown for the symmetry subspace $A_1$ (blue) and $^3\!A_2$ (orange) along with the spectrum for the lowest nine states. The bottom panel depicts the accuracy of the VQE results with respect to the exact diagonalization result in the respective symmetry subspace. The black lines represent the energy difference of the lowest excited states with respect to the ground state.}
\end{figure}

In this section the results for the VQE for the six-site molecule are shown in Fig.~\ref{fig:results_n6}. The six-site molecule was mapped to eight qubits using the Jordan-Wigner transformation optimised through tapering. To obtain the results, the in this paper presented variational form was used with a depth of $24$ CNOTs and optimized using up to $8000$ gradient descent steps for $10$ different CNOT configurations per $t'/t$. The initial states are choosen as the best performing parameters out of $20$ parameter sets after $2000$ optimization steps. The results are shown in Fig.~\ref{fig:results_n6}. We observe a sufficient accuracy to recapture the ground state transition showing that in principle the presented approach can also be applied for larger systems. The simulation on real quantum devices is not possible yet, as the error rates, especially the read out and entangling gate errors, are too large.

\subsection{All Results}\label{app:log_runs}
\begin{figure*}[t]
\includegraphics[scale=1]{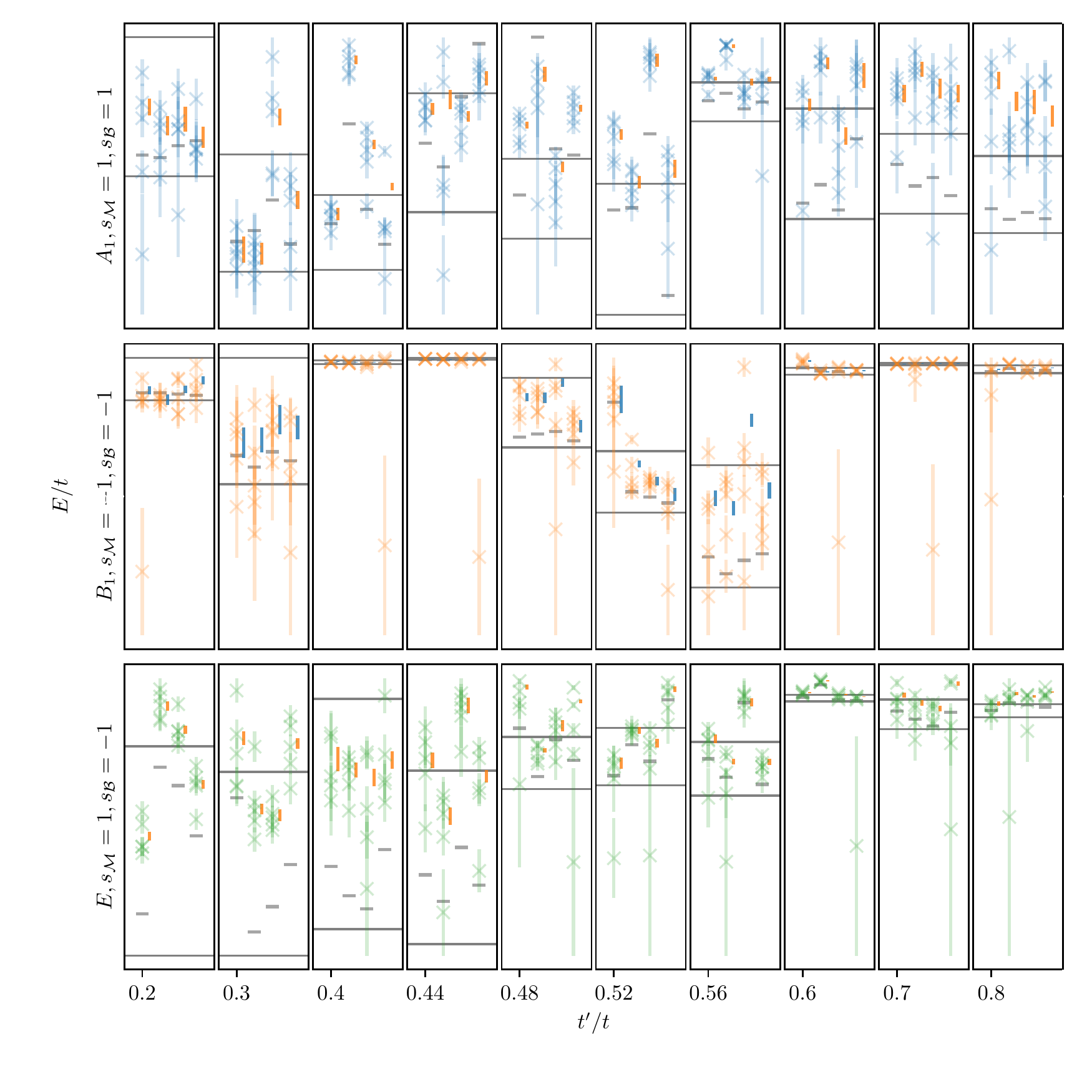}% Here is how to import EPS art
\caption{\label{fig:all_data}We show the results for the three discussed symmetry subspaces, $A_1$ (blue), $B_1$ (orange) and $E$ (green). For each symmetry subspace and each $t'/t$ we generate four entangling gate sequences, which are shown for different values at the $x$-axis for graphical reasons. The data within each panel belong to one $t'/t$. For each entangling gate sequences, we perform five measurements, which are shown in opaque colors. With the vertical lines, the result for the weighted least square is shown, as discussed in the main text. The small horizontal lines indicate the result, which we obtain by using the parameters $\bm \Theta$ found with the quantum computer and perform a non-noisy simulation. The two long horizontal lines indicate the energy scaling. The lower one corresponds to the exact diagonalisation value for the ground state energy $E_0$ in the respective subspace. The upper of the two long horizontal lines indicates $E_0 +0.1 t$.}
\end{figure*}

In this section we want to give all the results obtained on the quantum computer, of which a selection was shown in the main text. In Fig.~\ref{fig:all_data} all the measurement results for the Lanczos algorithm for all points are shown. Furthermore, the $y$ scale is chosen differently for each measurement results allowing to compare the results independently. As an extension to the main text, the Lanczos algorithm is also applied to the non-noisy simulations. This shows, that the Lanczos algorithm also provides a significant improvement without any noise. Furthemore, the observations explained in the main text can also seen for all data, so that the particular choice of data in the main text is representative.

\subsection{Convergence of the SPSA optimizer}
\begin{figure*}[t]
\includegraphics[scale=1]{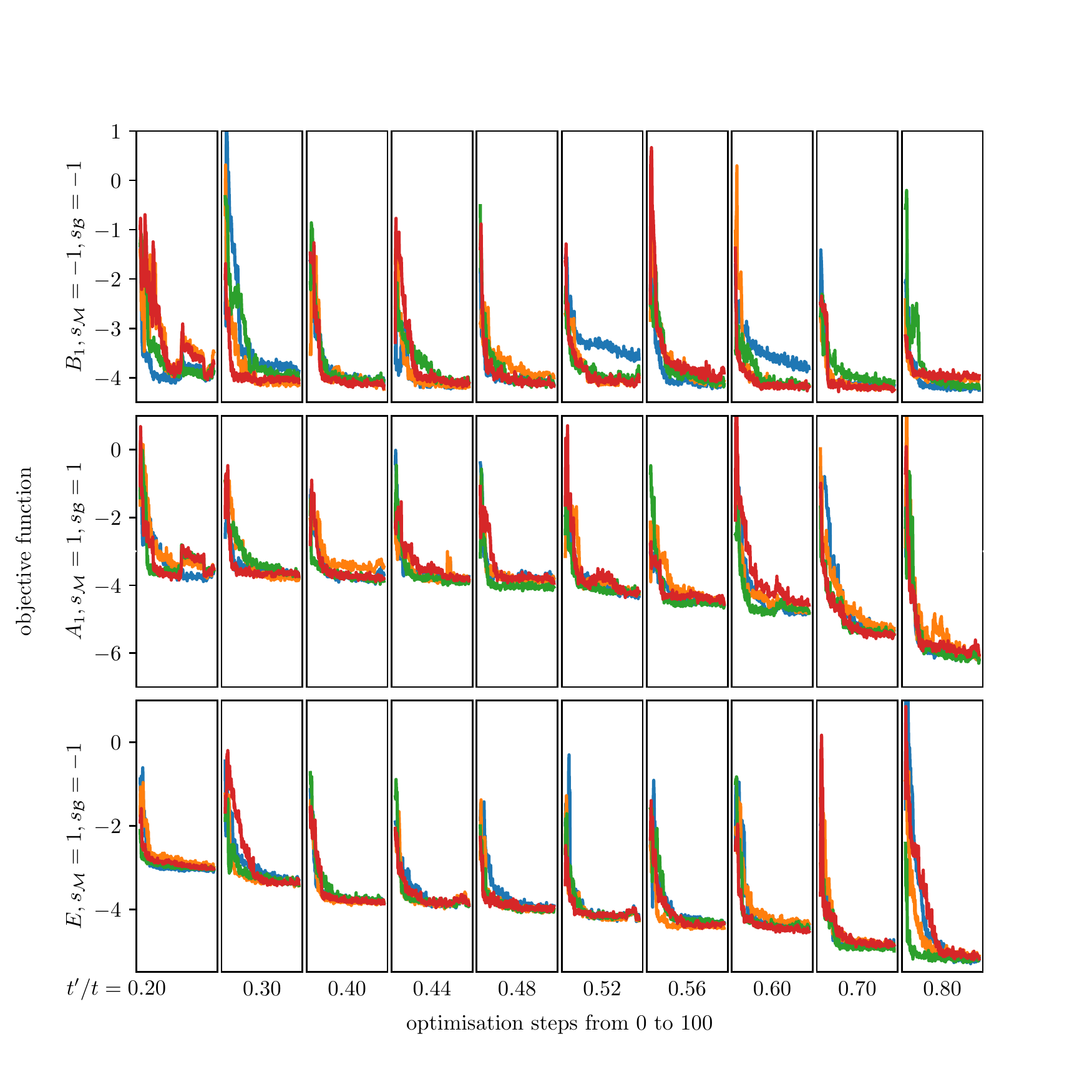}% Here is how to import EPS art
\caption{\label{fig:log_runs}
The optimization process for the data obtained on the quantum computer is shown. For each data point, given by $t'/t$ and its eigenvalue to the mirror symmetry  $s_\mathcal{M}$, the optimization process is shown. The rows correspond to results for the same symmetry subspace, in each column the results for a given $t'/t$ are shown. For each data point we choose four CZ sequences, so that in each panel four optimization processes are shown.}
\end{figure*}

In this section we show the results of the convergence procedure of the SPSA optimizer in Fig.~\ref{fig:log_runs}. They indicate, that the selected number of $100$ optimization steps was sufficient in most cases. For the symmetry subspace belonging to $A_1$ for high $t'/t$ convergence was not reached. Therefore, the results might be improvable by performing more steps. However, this improvement is not expected to be significant.

\section{Acknowledgements.}
Authors acknowledge interesting discussions with Pauline J. Ollitrault and Igor Sokolov. 

This project has received funding from the European Research Council (ERC) under the European Union’s Horizon 2020 research and innovation programm (ERC-StG-Neupert-757867-PARATOP). 
IT acknowledges the financial support from the Swiss National Science Foundation (SNF) through the grant No. 200021-179312. IBM, the IBM logo, and ibm.com are trademarks of International Business Machines Corp., registered in many jurisdictions worldwide. Other product and service names might be trademarks of IBM or other companies. The current list of IBM trademarks is available at \url{https://www.ibm.com/legal/copytrade}.

% \section{Author contributions}
% T.N. conceived the idea of simulating the ring-like Hubbard molecule with a quantum computer. All authors contributed to the development of the present approach. P.S. performed all simulations and data analysis. All authors discussed the results and wrote the manuscript.

% \section{Competing interests}
% The authors declare no competing interests.

%\nocite{*} %only the last citation is not included yet, maybe I need it for the 6 side system

%

%

\end{document}